\documentclass[twoside,leqno]{article}
\usepackage{float}
\usepackage[letterpaper]{geometry}

\usepackage{siamproceedings}

\usepackage[T1]{fontenc}
\usepackage{amsfonts}
\usepackage{graphicx}
\usepackage[numbers]{natbib}
\usepackage{epstopdf}
\usepackage{enumitem}
\usepackage{algorithm}
\usepackage{algpseudocode}
\ifpdf
  \DeclareGraphicsExtensions{.eps,.pdf,.png,.jpg}
\else
  \DeclareGraphicsExtensions{.eps}
\fi

\newsiamremark{remark}{Remark}
\newsiamremark{hypothesis}{Hypothesis}
\crefname{hypothesis}{Hypothesis}{Hypotheses}
\newsiamthm{claim}{Claim}

\usepackage{amsopn}

\begin{document}

\newcommand\relatedversion{}

\title{\Large Towards bridging the gap between data-driven and theoretical turbulence closures in stratified flows \relatedversion}
    \author{Laure Zanna\thanks{Courant Institute School of Mathematics, Computing and Data Science, New York University, New York, NY, USA, Center for Data Science, New York University, New York, NY, USA  (\email{laure.zanna@nyu.edu}).} 
    \and Pavel Perezhogin\thanks{Courant Institute School of Mathematics, Computing and Data Science, New York University, New York, NY, USA, Center for Data Science, New York University, New York, NY, USA (\email{pp2681@nyu.edu})}} 
    
\date{}

\maketitle


\fancyfoot[R]{\scriptsize{Copyright \textcopyright\ 20XX by SIAM\\
Unauthorized reproduction of this article is prohibited}}





\begin{abstract}

Turbulence closure models are essential for solving the equations of motion in realistic systems, where fully resolving all relevant scales of motion is computationally infeasible. 
Developing turbulence closures remains one of the most challenging problems in fluid dynamics.
Specifically, the Navier–Stokes equations, when filtered to isolate large-scale motions, introduce new terms representing the influence of subgrid-scale turbulent stresses.  
These terms, which can only be computed directly by resolving the turbulence itself, therefore lead to the closure problem:  we must add new equations or introduce assumptions to relate the unresolved scales of motions to the resolved flow. Here we consider the closure problem for oceanic flows, i.e., stratified, Boussinesq, incompressible, in a rotating frame of reference. In particular, we focus on a closure for ocean mesoscale eddies, which have horizontal scales of 10-100km and are key to the redistribution of momentum, energy, and tracers in the ocean. In particular, mesoscale eddies can reinject energy and momentum into the large-scale flow through an inverse energy cascade.
Here, we explore a range of theoretical and data-driven ocean mesoscale closures and examine their connections using analytical and data-driven methods. 
This note aims to bridge the gap between novel methods from artificial intelligence (AI) and machine learning and theoretical fluid dynamics to address significant challenges in the physics of turbulence. 
\end{abstract}

\newpage

\section{Introduction.}

Turbulence is one of the most fundamental and ubiquitous phenomena in fluid dynamics, including in geophysical flows such as atmospheric storms and ocean currents. 
For example, turbulent motions can influence the transport of momentum, energy, heat, and tracers in fluid flows. 
Energy is injected, redistributed, and dissipated across a wide range of scales. 
Understanding the energy cycle is essential for predicting and modeling large-scale phenomena such as weather patterns, ocean circulation, and atmospheric jets. 

In three-dimensional turbulence, energy typically cascades from large scales—where it is injected by external forcing, e.g., winds - down to smaller scales where viscosity ultimately dissipates it. 
However, in two-dimensional (2D) or quasi-two-dimensional flows—such as large-scale atmospheric and oceanic motions constrained by stratification and rotation—the dynamics are markedly different. 
Rather than cascading energy downward, turbulent eddies transfer kinetic energy upscale through an inverse energy cascade, creating larger flow structures. 
However, enstrophy (defined as vorticity squared) cascades downscale, leading to dissipation at small scales. 
This dual cascade of energy and enstrophy in 2D stratified fluid leads to the spontaneous emergence of coherent, long-lived vortices and jets. 
These vortices or eddy motions are critical for maintaining large-scale ocean currents such as the Gulf Stream, the Kuroshio, and the Southern Ocean Antarctic Circumpolar Current. 

Numerical ocean models, which solve the equations of motion for the ocean on a grid, often exhibit weak kinetic energy (KE) and reduced effective tracer diffusivity compared to high-resolution simulations. 
These errors (or biases) are often due to the coarse resolution of the model grid, such that the turbulence is largely unresolved and
excessive numerical viscosity is employed for model stability. 
The biases can be reduced by improving the
subgrid closures (or parametrization) of unresolved motions and their sub-grid statistics \cite{Brankart2013,Jansen2014,Bachman2017,perezhogin_2025_dataset}. 

Several closures have been proposed and implemented in a range of numerical simulations, with varying degrees of complexity and success.  
Here, we analyze a proposed closure, based on the Rivlin-Ericksen \cite[][hereafter RE]{Rivlin1955} stress tensor. 
A version of the RE stress tensor has been successfully implemented as a parametrization in quasi-geostrophic models \cite{Zanna2017}. 
\cite{PortaMana2014} and \cite{Zanna2017} found that the approximate RE parametrization used could mimic unresolved momentum fluxes and enhance eddy-mean flow interactions, improving the energy cycle of the simulation. 
\cite{Anstey2017} derived another form of the RE parametrization based on primitive equations in an idealized ocean general circulation model. 
\cite{Anstey2017} presented the RE parametrization as being comprised of both a "deformation" term and a "memory" term, where the "deformation" term only depends on the current spatial derivatives of velocity, while the "memory" term depends on both the current and previous spatial derivatives of velocity. 
They showed that the "deformation" term can redistribute KE but not add to the global-mean KE, while dissipating enstrophy, and that it may be a good candidate for parametrizing unresolved momentum fluxes in ocean models. 
Later, \cite{Zanna2020} and \cite{perezhogin_2025_dataset} demonstrated that the deformation-based closure of \cite{Anstey2017} shares similarities with nonlinear gradient models under certain assumptions. The closure has been successfully implemented in idealized and global models \cite{Zanna2020,perezhogin_2025_dataset}.
The RE parametrization "memory" term has not yet been explored for implementation in any primitive-equation ocean general circulation. 
However, the full RE closure, i.e., both the "deformation" and "memory" terms, shares some similarities with the LANS-$\alpha$ parametrization \cite{Holm2005,Bachman2018}, which has been implemented in an ocean general circulation model \cite{Hecht2008} and found to reduce biases in turbulent regions. 
  
In this paper, we further explore the potential closures for ocean turbulence, using machine learning to extract an equation from high-resolution data. We then compare the discovered closure and its properties to the RE closures, including the RE-memory part, which was previously ignored by \cite{Anstey2017}. 

\section{Turbulence closures in a stratified fluid.}
In this section, we introduce the problem of parameterization of subgrid eddies in the governing equations of ocean dynamics. For the typical horizontal resolution of global ocean models, the dominant missing turbulent forcing is due to mesoscale eddies.
Ocean mesoscale eddies are commonly defined as coherent structures and filaments with horizontal scales on the order of the first baroclinic deformation radius, $L_d$, which characterizes the natural length scale of stratified, rotating flows and is given by
$
L_d = \frac{c}{f},
\quad \text{with} \quad
c = \sqrt{g' H},
$
Here $c$ is the phase speed of the first baroclinic mode, $g' = g(\Delta \rho / \rho_0)$ is the reduced gravity representing stratification with $\Delta \rho$ - the vertical density difference, $\rho_0$ a reference density, and $H$ is the upper layer thickness. In addition, $f = 2\Omega \sin\phi$ is the Coriolis parameter, with $\Omega$ Earth’s rotation rate and $\phi$ latitude. Mesoscale eddies typically have horizontal scales  $R \sim L_d$, which range from $\sim 10$  km at high latitudes to $\sim 100$ km in the tropics \cite{Hallberg2013}. They are characterized by small Rossby numbers $ |\zeta/f| \ll 1$, where the relative vorticity $\zeta= \frac{\partial v}{\partial x} - \frac{\partial u}{\partial y}$, and with dynamics governed by a balance between pressure gradients and Coriolis forces. 

\subsection{Momentum equations.} Consider the standard approximations for oceanic flows, that is, the fluid is stratified, incompressible, and Boussinesq fluid in a rotating frame of reference. In addition, we assume that the hydrostatic balance holds (i.e., the vertical pressure gradient force balances the gravity acceleration). 

Let us define the velocity vector of the fluid, $\mathbf{u} = \left(u,v,w\right) = \mathbf{u}_{h} + w \, \mathbf{k} $, where $u$, $v$, $w$ are the zonal, meridional, vertical component of the velocity, the subscript ${h}$ denotes the vector component in the horizontal plane, and the unit vector $\mathbf{k}$ points in the vertical direction.

The momentum balance for the horizontal velocity vector for such fluid is given by 
\begin{equation}
    \partial_{t} \, \mathbf{u}_{h}
    + \left[ (\mathbf{u} \cdot  \mathbf{\nabla})   \mathbf{u} \right]_{h}
    + f \, \mathbf{k} \times \mathbf{u}_{h}
    + \frac{1}{\rho_{0}} \, \mathbf{\nabla}_{h} p = \mathbf{F} + \mathbf{D}.
    \label{eq:momentum_equation_2}
\end{equation}
Here $p$ is the pressure,  $f$ is the Coriolis frequency, $\rho_{0}$ is a reference density, $\mathbf{F}$ is the external forcing, i.e., wind stress forcing at the ocean surface and friction along the solid boundaries, and $\mathbf{D}$ is the viscosity. 

\subsection{Lateral momentum closure.}  In ocean numerical simulations, the governing equations are discretized on a spatial grid with horizontal resolution far larger than the length scale at which molecular viscosity acts on the flow. 
The viscous stresses, captured by $\mathbf{D}$, are therefore a combination of unresolved turbulent processes and numerical effects to keep the simulations numerically stable. 
The unresolved turbulent processes are captured using a closure model (also called a parametrization), which represents the effect of fine-scale processes on the large-scale flow. 

To define the closure problem, we introduce a filtering operator $\overline{\left(\cdot\right)}$, which maps fine-grid variables onto a coarser grid. The filtered horizontal momentum equation is then defined by
\begin{equation}
    \partial_{t} \, \overline{\mathbf{u}}_{h}
    + \left[ (\overline{\mathbf{u}} \cdot  \overline{\mathbf{\nabla}})   \overline{\mathbf{u}} \right]_{h}
    + f \, \mathbf{k} \times \overline{\mathbf{u}}_{h}
    + \frac{1}{\rho_{0}} \, \mathbf{\nabla}_{h} \overline{p} = \mathbf{\overline{F}} + \mathbf{\overline{D}} + \mathcal{S}.
    \label{eq:momentum_coarsegrained}
\end{equation}
The form of the filtered equation is equivalent to the unfiltered \autoref{eq:momentum_equation_2}, with an additional forcing term, $\mathbf{\mathcal{S}}$, which represents the effect of nonlinear advection terms on the filtered (coarse-resolution) flow and is given by  \begin{equation}
    \mathbf{\mathcal{S}}  = \left[ (\overline{\mathbf{u}} \cdot  \overline{\mathbf{\nabla}})   \overline{\mathbf{u}} \right]_{h} - \overline{\left[ (\mathbf{u} \cdot  \mathbf{\nabla})   \mathbf{u} \right]}_{h}.     \label{eq:subgrid_mom_flux} 
\end{equation}

The closure problem is to attempt to close equation \ref{eq:momentum_coarsegrained}, that is, to propose a function that predicts the subgrid forcing given only the information about the large-scale flow: $ \mathbf{\mathcal{\hat{S}}} (\overline{\mathbf{u}})\approx  \mathbf{\mathcal{S}} (\overline{\mathbf{u}}, \mathbf{u})$. Below, we consider only horizontal velocities and thus omit the index $h$ for brevity.
\\

\section{Data-driven lateral momentum closure}
In this section, we address the turbulence closure problem using data-driven techniques that leverage high-resolution simulation data and machine learning methods.
\subsection{Diagnostics.} To diagnose the subgrid forcing from data, we use a dataset from a high-resolution model and 
follow the procedure described in \cite{perezhogin2025generalizable}.

The dataset is derived from the filtered output of the climate model CM2.6 \cite{griffies2015impacts}, which has a horizontal resolution of $0.1^\circ$.
In practice, the filtering operation is performed by applying a Gaussian spatial filter to suppress small-scale variability, followed by a coarse-graining operation (weighted by the local grid cell area). 
The filtering step before coarse-graining is necessary when there is not enough scale separation between the high-resolution and coarse-resolution grids to remove the ringing effect from discretization errors \cite{guillaumin2021stochastic,christensen2022parametrization}.

We consider the filtering and coarse-graining towards a target of $0.4^\circ$ horizontal resolution, which is typically referred to as eddy-permitting.
In this regime, mesoscale turbulence, which is determined by the ocean stratification and the frequency of rotation, is partially but not fully resolved. 

\begin{figure}
    \centering
    \includegraphics[width=1.0\linewidth]{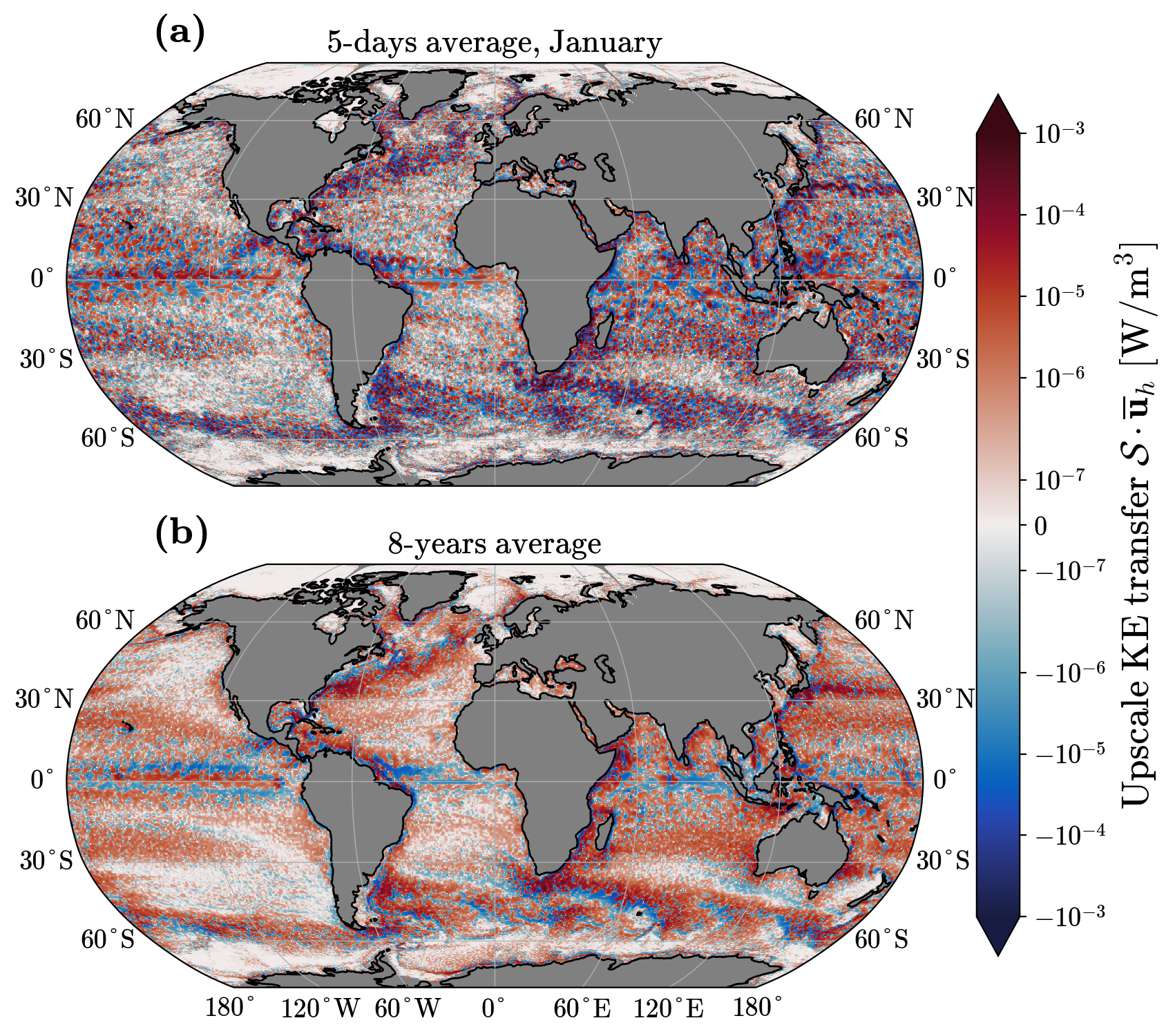}
    \caption{Upscale kinetic energy (KE) transfer diagnosed in  the state-of-the-art climate model CM2.6 \cite{griffies2015impacts}, here shown for a subsurface ocean layer at 5m depth. The horizontal resolution of the ocean component is $1/10^\circ$, the horizontal resolution of the coarse grid is $0.4^\circ$, and the filter scale is $1.2^\circ$, see \cite{perezhogin2025generalizable}. Panel (a) shows a snapshot of the KE transfer (averaged over the time sampling interval, 5 days), and panel (b) shows the 8-year average of the KE transfer.}
    \label{fig:KE_transfer}
\end{figure}

In the eddy-permitting regime, the net effect of $\mathcal{S}$ is an upscale kinetic energy transfer from unresolved subgrid eddies to the resolved eddy field ($\mathcal{S} \cdot \overline{\mathbf{u}}_{h} > 0$, see Figure \ref{fig:KE_transfer}(b)), typical for quasi-2D stratified fluid. 
However, locally, the transfer of KE is not sign definite and depends on the local velocity, as well as its temporal and spatial gradients, see Figure \ref{fig:KE_transfer}(a).

\subsection{Equation discovery.}

To distill the information contained in the subgrid forcing and determine an appropriate closure for ocean mesoscale turbulence, we follow the procedure described in ~\cite{koza1994genetic,ross2022benchmarking}. 
Specifically, we learn an equation from the diagnosed subgrid data and compare it to previously proposed theoretical closures.  

The method proposed by \cite{ross2022benchmarking} combines sparse linear regression given a pre-defined feature library \cite{brunton2016sparse, li2021,Zanna2020} and genetic programming (GP) ~\cite{10.1093/mind/LIX.236.433,koza1994genetic}. 
GP algorithms do not rely on an explicit library of functions but rather on a set of atomic features
and a set of operations for combining them. The algorithm \ref{alg:cap} from \cite{ross2022benchmarking} combines both methods to ensure more robust learned expressions, while respecting physical units. 

\begin{algorithm}
\begin{algorithmic}[1]
\Procedure{FitGeneticProgram}{$x$, $y$}
\State Run \texttt{gplearn} \cite{gplearn} with operators $\{\partial_x,\partial_y,\nabla^2,(\mathbf{\overline{u}}\cdot \nabla),*,+\}$, and $$\mathrm{Fitness}(\mathtt{term}) = |\mathrm{Corr}(\mathtt{term}(x), y)| -0.001 * \mathrm{Length}(\mathtt{term})$$
\EndProcedure
\\
\Procedure{FitLinearRegression}{$x$, $y$}
\State Find $w$ to minimize $||w \cdot x - y||_2^2$
\EndProcedure
\\
\Procedure{FitHybridSymbolic}{$x$, $y$}
\State $\mathtt{terms} \gets \emptyset$
\Comment set of symbolic expressions
\State $w \gets \emptyset$
\Comment weights of those expressions
\State $\tilde{y} \gets$ y
\Comment residual forcing to predict
\\
\Repeat
    \ForAll{layers $z$}
        \State $\mathtt{terms} \gets \mathtt{terms}\, \cup$ \Call{FitGeneticProgram}{$x_z, \tilde{y}_z$}
        \Comment learn the next term
    \EndFor
    \State {$\mathtt{terms} \gets$ \Call{OptionalUserEdits}{$\mathtt{terms}$}}
    \ForAll{layers $z$}
        \State $w_z \gets$ \Call{FitLinearRegression}{$\mathtt{terms}(x_z), y_z$}
        \Comment reweight terms
        \State $\tilde{y}_z \gets w_z \cdot \mathtt{terms}(x_z) - y_z$ 
        \Comment update residuals
    \EndFor
\Until{convergence {or user decision}}
\\
\State \textbf{return} \texttt{terms}, $w$
\EndProcedure
\end{algorithmic}
\caption{``Hybrid'' linear and genetic programming-based symbolic regression (with optional human-in-the-loop interventions). Code reproduced from \cite{ross2022benchmarking}.}\label{alg:cap}
\end{algorithm}

By applying the hybrid equation-discovery method, we extract the following expression 
\begin{equation}
\mathcal{S} = 
\alpha_1 \left( \nabla \mathbf{u} \nabla \right)^T \nabla \mathbf{u}^T  
+ \alpha_2  \left( \nabla \mathbf{u}^T \nabla^2  \mathbf{u}\right), 
\label{eqn:discovered_equation}
\end{equation}
the parameters $\alpha_1$ and $\alpha_2$ are the linear weights found. 
This expression is robust; i.e., it is insensitive to the initialization of the algorithm, as well as to the algorithm itself (e.g., when using simple regression versus genetic programming). Other terms were also found but did not yield the same robustness as the ones described in \ref{eqn:discovered_equation}. 
We will revisit this expression later in the manuscript, after we have explored the potential theoretically motivated turbulence closures. 



\section{Theory}
In this section, we draw connections between the identified data-driven closure \eqref{eqn:discovered_equation} and existing parametrizations.


\subsection{The Rivlin Eriksen-based closure.}

We use a form of the two-dimensional (2D) parametrization proposed by \cite{PortaMana2014}, \cite{Anstey2017} and \cite{Zanna2017}, originally based on the Rivlin Eriksen (RE) stress tensor for non-Newtonian fluids. 
We follow \cite{Anstey2017}, who presented the RE stress tensor for primitive-equation models, and define the 2D velocity gradient tensor as
\begin{equation}
\mathbf{\nabla u}_h = \left[ 
{\begin{array}{cc}
   u_x & u_y \\
   v_x & v_y \\
  \end{array} } \right], \nonumber
\end{equation}
where $u$ and $v$ are zonal and meridional velocities and subscripts denote derivatives in the zonal, $x$, and meridional, $y$, directions, respectively. 
The 2D gradient operator is $\nabla_h = (\partial / \partial_x, \partial / \partial_y)$. 
For ease of notation, we will drop the subscript $h$ in this section and for the rest of the paper. 

The 2D velocity gradient tensor can be split into symmetric and anti-symmetric parts, $\mathbf{S} = \frac{1}{2} (\mathbf{\nabla u} + \mathbf{\nabla u}^T)$ and $\mathbf{W} = \frac{1}{2} (\mathbf{\nabla u} - \mathbf{\nabla u}^T)$, which are the strain and vorticity tensors, respectively. 
The RE stress tensors for non-Newtonian fluids \cite{Rivlin1955} in two dimensions are given by 
\begin{eqnarray}
\mathbf{A}_1 &=& 2 \mathbf{S}, \label{eq:a1} \\
\mathbf{A}_2 &=& \frac{D \mathbf{A}_1}{Dt} + \mathbf{\nabla u}^T \mathbf{A}_1 + \mathbf{A}_1 \mathbf{\nabla u}, \label{eq:a2} \\ 
&& \dots \nonumber \\
\mathbf{A}_m &=& \frac{D \mathbf{A}_{m-1}}{Dt} + \mathbf{\nabla u}^T \mathbf{A}_{m-1} + \mathbf{A}_{m-1} \mathbf{\nabla u}, \nonumber
\end{eqnarray}
where the Lagrangian derivative, $D / Dt = \partial/\partial t + u \partial / \partial x + v \partial / \partial y$, is the time derivative following a fluid parcel and the subscript $m$ denotes the increasing order of the stress tensors. 
The momentum tendency associated with the RE stress tensor is the 2D divergence of the stress tensor,  $\nabla \cdot \mathbf{A}_m$. 

We assume that the 2D flow is non-divergent, i.e., $\nabla \cdot \mathbf{u} = 0$, which simplifies the derivation and is supported by observed and simulated small vertical velocities in the stratified ocean. 
Given $\nabla \cdot \mathbf{u} = 0$, the momentum forcing when $m=1$ (Eq. \ref{eq:a1}) is given by $\nabla \cdot (\nu 2 \mathbf{S}) = \nabla  \cdot \nu\nabla \mathbf{u}$, i.e. Laplacian viscosity, where $\nu=\mathrm{const}$ is the viscosity coefficient \citep{Anstey2017}. 
Parametrizing subgrid-scale motions as Laplacian or bi-Laplacian Smagorinsky viscosity is commonly done in ocean models \citep{Griffies2000}. However, 
\cite{Anstey2017} proposed a parametrization based on $m=2$ (Eq. \ref{eq:a2}) and diagnosed the deformation term, $\mathbf{\nabla u}^T \mathbf{A}_1 + \mathbf{A}_1 \mathbf{\nabla u}$, which we will hereafter denote as the RE-AZ (or RE-deformation) parametrization.  

\subsubsection{The RE deformation-based closure.}
We consider here the part of $\mathbf{A}_2$ that does not involve a time derivative, the term $\nabla \mathbf{u} ^\mathrm{T} \mathbf{A}_1  +  \mathbf{A}_1 \nabla \mathbf{u}$, which is a function of the instantaneous state of the flow. Using $\nabla \mathbf{u} = \mathbf{S} + \mathbf{W}$   and  $\nabla \mathbf{u}{^\mathrm{T}} = \mathbf{S} - \mathbf{W} $ and following \cite{Anstey2017}, the RE-AZ parameterization is given by
\begin{equation}
    \mathbf{\nabla u}^T \mathbf{A}_1 + \mathbf{A}_1 \mathbf{\nabla u}  =  4\mathbf{S}^2  + 2 (\mathbf{S}\mathbf{W}-\mathbf{W}\mathbf{S}). \label{eqn:AZ2017}
\end{equation}

The RE-AZ parametrization was found to redistribute KE while conserving domain-averaged KE and dissipating enstrophy. 

\subsubsection{The RE memory-based closure.}
Here we will diagnose and test the part of Eq. \ref{eq:a2} that RE-AZ mostly ignored, namely the memory term $\frac{D \mathbf{A}_\mathrm{1}}{Dt}$, and evaluate its suitability as a parameterization. 
We will call this the "RE-memory" parameterization, since the term $D \mathbf{A}_1 / Dt$  represents the effects of the parcel's history of deformation on the present eddy stress. 
We will show that such a parameterization has the potential to inject KE and potentially alleviate the lack of KE found in eddy-permitting ocean models.

The RE-memory stress tensor for $m=2$ can be expanded as 
\begin{eqnarray}
\frac{D \mathbf{A}_1}{Dt} &=& 2  \frac{d \mathbf{S}}{dt} = 2\frac{D}{Dt} (\mathbf{\nabla u} - \mathbf{W})  = \nonumber \\
&=&  -2  \frac{D \mathbf{W}}{Dt} + 2 \frac{D \mathbf{\nabla u}}{Dt} = \nonumber \\
&=& -2  \frac{D \mathbf{W}}{Dt} + 2  \nabla \frac{D \mathbf{u}}{Dt} - 2  \nabla \mathbf{u} \nabla \mathbf{u}. \label{eq:re_m2}
\end{eqnarray}
We ignore the term $D \mathbf{W} / Dt$, as no eddy stress is generated for purely rotational flows \citep{Wajsowicz1993,Smith2002}. 
If we assume a non-divergent flow, we can write
\begin{equation}
2 \nabla \mathbf{u} \nabla \mathbf{u} = 2\left(\mathbf{S}^2 + \mathbf{W}^2 \right) = \frac{1}{2} (\delta^2 - \zeta^2) \mathbf{I}, \label{eqn:trace}
\end{equation}
where $\mathbf{I}$ is the identity matrix, $\zeta = v_x - u_y$ is the vorticity, and  $\delta = \sqrt{\tilde{D}^2 + D^2 }$ is the total deformation. 
Stretching deformation and shearing deformation are denoted by $\tilde{D} = u_x - v_y$ and $D = u_y + v_x$, respectively, which represent changes in the shape of a fluid parcel.
Note also that the Smagorinsky \cite{smagorinsky1963general} viscosity coefficient depends on the flow deformation.

The momentum forcing of the RE-memory parameterization is thus given by the divergence of the second and third terms in Eq. \ref{eq:re_m2}, 
\begin{eqnarray}
\nabla \cdot \kappa \frac{D \mathbf{A}_1}{Dt} &=& \nabla \cdot \left( 2 \kappa \nabla \frac{D \mathbf{u}}{Dt} - 2\kappa  \nabla \mathbf{u} \nabla \mathbf{u} \right) = 2 \kappa \nabla^2 \frac{D \mathbf{u}}{Dt} - \frac{1}{2} \kappa \nabla \cdot (\delta^2 - \zeta^2) \mathbf{I} \label{eq:re_eul}\\
&=&  2 \kappa \frac{D \nabla^2 \mathbf{u}}{Dt} + 2 \kappa (\nabla^2 \mathbf{u} \cdot \nabla) \mathbf{u} - \frac{1}{2} \kappa \nabla \cdot (\delta^2 - \zeta^2) \mathbf{I}, \label{eq:re_lagr}
\end{eqnarray}
where $\kappa$ is assumed here to be a constant coefficient and $\nabla^2 \mathbf{u} = (\nabla^2 u, \nabla^2 v)$. 
\cite{PortaMana2014} and \cite{Zanna2017} found that $\kappa$ can be set as 
\begin{equation}
\kappa = \gamma (\Delta x)^2, 
\end{equation}
where $\gamma$ is a constant, i.e., we assume that spatial and temporal variations in the RE parametrization arise only from the dependency on $\mathbf{S}$ and $\mathbf{W}$ and not from the parameters themselves.
In an ocean model using a longitude-latitude grid, $\Delta x$ is typically spatially varying. 
\cite{Zanna2017} used $\gamma \approx -(0.3)^2$ and found that their form of the RE parametrization injected KE and amplified the inverse cascade of KE. Choosing the correct sign ($\gamma<0$ and $\kappa<0$) is important to parameterize injection, but not dissipation of KE. We note that \cite{PortaMana2014} and \cite{Zanna2017} employed a quasi-geostrophic model, and thus their form of the RE-memory parametrization differed. 
\cite{Bachman2018} noted that using $\nabla \cdot \kappa \mathbf{A}_2$ to parametrize subgrid-scale momentum fluxes bears some mathematical similarities to the LANS-$\alpha$ parametrization \citep{Holm2005,Hecht2008}, in which case the coefficient $\kappa$ is similar to their $-\alpha^2$. 


For an implementation in a primitive-equation ocean model, it is attractive to write the RE-memory parametrization as in Eq. \ref{eq:re_eul} 
\begin{equation}
\frac{\partial \mathbf{u}}{\partial t} = 2 \kappa \nabla^2 \frac{D \mathbf{u}}{Dt} - \frac{1}{2} \kappa \nabla \cdot (\delta^2 - \zeta^2) \mathbf{I},
\end{equation}
since $D \mathbf{u}/Dt$ is readily available in the model. We will return to terms in Eq. \ref{eq:re_eul} later in this section. 



\subsection{Effects on kinetic energy}

We now turn to the effect of the RE-memory parametrization on the global 2D KE budget. 
In two dimensions, the KE tendency from the RE-memory parametrization can be found by multiplying the momentum tendency (Eq. \eqref{eq:re_eul}) by the velocity,
\begin{eqnarray}
\label{eq:ure_lagr}
\frac{\partial E_K}{\partial t} = \mathbf{u} \cdot \kappa \nabla^2 \frac{\partial \mathbf{u}}{\partial t} + \mathbf{u} \cdot  \kappa \nabla^2 \left[ (\mathbf{u} \cdot \nabla) \mathbf{u} \right] - 
\mathbf{u} \cdot \frac{1}{2} \kappa \nabla \cdot (\delta^2 - \zeta^2) \mathbf{I}. 
\end{eqnarray}

We note that assuming $\nabla \cdot \mathbf{u}=0$, 
\begin{eqnarray}
\mathbf{u} \cdot \kappa \nabla^2 \frac{\partial \mathbf{u}}{\partial t} &=& \nabla \cdot \left( \mathbf{u} \cdot \kappa \nabla \frac{\partial \mathbf{u}}{\partial t} \right) - \nabla \mathbf{u} : \kappa \nabla \frac{\partial \mathbf{u}}{\partial t}, \nonumber \\
\mathbf{u} \cdot  \kappa \nabla^2 \left[ (\mathbf{u} \cdot \nabla) \mathbf{u} \right] &=& \nabla \cdot \left( \mathbf{u} \cdot \kappa \nabla \left[ (\mathbf{u} \cdot \nabla) \mathbf{u} \right] \right) - \nabla \mathbf{u} : \kappa \nabla \left[ (\mathbf{u} \cdot \nabla) \mathbf{u} \right] \nonumber \\
\mathbf{u} \cdot \frac{1}{2} \kappa\nabla \cdot (\delta^2 - \zeta^2) \mathbf{I}&=& \nabla \cdot \left(\frac{1}{2}\kappa\mathbf{u} (\delta^2-\zeta^2)\right), \nonumber 
\end{eqnarray}
where $(:)$ is the tensor contraction over two indices.

\begin{figure}[h]
\centerline{\includegraphics[width=0.7\linewidth]{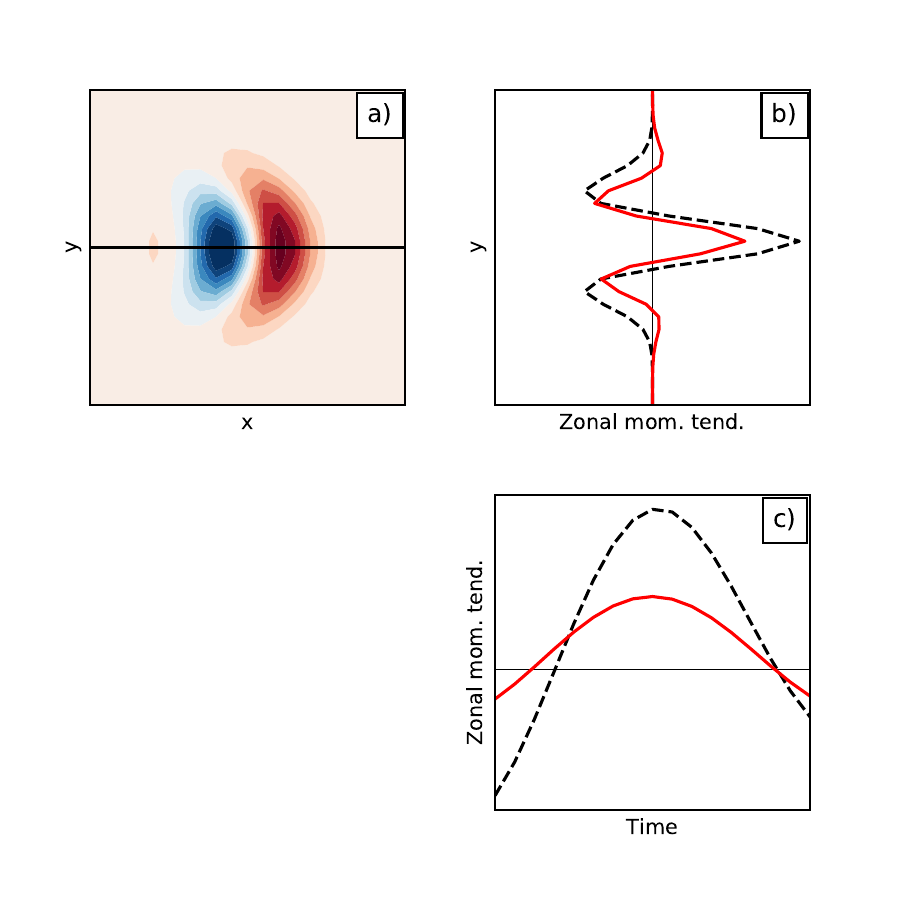}}
\caption{
a) Schematic eddy deformation similar to \cite{Anstey2017}. 
b) Zonal mean zonal momentum tendency from the flow (black dashed) and the advective part of the parametrization (red). 
c) Zonal mean zonal momentum tendency in the center of the domain (black solid line in panel (a)) from the local time derivative of the flow (black dashed) and the parametrization (red). 
}
\label{fig:james_eddy}
\end{figure}

We denote a global area integral as $\lbrace \mathbf{\phi} \rbrace = \int \int \mathbf{\phi} ~dx ~dy$, and find that 
\begin{eqnarray}
\left\lbrace \frac{\partial E_K}{\partial t} \right\rbrace &=& \left\lbrace \nabla \cdot \left( \mathbf{u} \cdot \kappa \nabla \frac{\partial \mathbf{u}}{\partial t} \right) \right\rbrace - \left\lbrace \nabla \mathbf{u} : \kappa \nabla \frac{\partial \mathbf{u}}{\partial t} \right\rbrace + \nonumber \\
&+& \left\lbrace \nabla \cdot \left( \mathbf{u} \cdot \kappa \nabla \left[ (\mathbf{u} \cdot \nabla) \mathbf{u} \right] \right) \right\rbrace - \left\lbrace \nabla \mathbf{u} : \kappa \nabla \left[ (\mathbf{u} \cdot \nabla) \mathbf{u} \right] \right\rbrace +\left\lbrace\nabla \cdot \left(\frac{1}{2}\kappa\mathbf{u} (\delta^2-\zeta^2)\right) \right\rbrace
\nonumber  \\
&=& - \left\lbrace \nabla \mathbf{u} :\kappa \nabla \frac{\partial \mathbf{u}}{\partial t} \right\rbrace - \left\lbrace \nabla \mathbf{u} : \kappa \nabla \left[ (\mathbf{u} \cdot \nabla) \mathbf{u} \right] \right\rbrace \nonumber\\ 
&=& - \left\lbrace \frac{1}{2} \kappa \frac{\partial}{\partial t} \left( \left| \nabla \mathbf{u} \right|^2 \right) \right\rbrace - \left\lbrace \nabla \mathbf{u} : \kappa \nabla \left[ (\mathbf{u} \cdot \nabla) \mathbf{u} \right] \right\rbrace \label{eq:ke_tend}, 
\end{eqnarray}
since the RE-memory parametrization is only evaluated in 2D, the global integral of a divergence vanishes. 
%
Eq. \ref{eq:ke_tend} shows that the global KE tendency depends on the global integral of the tendency in absolute velocity gradient 
and the relative orientation of the velocity gradient to the gradient of momentum advection. 
For example, if the tendency $\partial \left|\nabla \mathbf{u}\right|^2 / \partial t$ generally is positive, i.e., velocity gradients are sharpening, then 
$-\left\lbrace \frac{1}{2} \frac{\partial}{\partial t} \left( \left| \nabla \mathbf{u} \right|^2 \right) \right\rbrace < 0$. 
Furthermore, if momentum advection generally acts to sharpen velocity gradients, e.g., transfer momentum from the flanks to the core of a jet, then $\left\lbrace \mathbf{u} \cdot (\mathbf{u} \cdot \nabla) \mathbf{u} \right\rbrace >0$ and $- \left\lbrace \nabla \mathbf{u} :   \nabla \left[(\mathbf{u} \cdot \nabla) \mathbf{u}\right] \right\rbrace < 0$. 
Hence, if $\gamma < 0 \Rightarrow \kappa < 0$, the first and second terms on the right-hand side in Eq. \ref{eq:ke_tend} are positive and amplify the globally integrated KE tendency. 
This is illustrated in Fig. \ref{fig:james_eddy}, where we show a highly idealized eddy that transfers momentum towards a jet core, and the momentum transfer and its temporal variability are enhanced by the RE-memory parametrization when $\gamma < 0$.

\subsection{Effects on eddy phase speed}










In a two-dimensional, non-divergent flow, the RE-memory parametrization contributes to the vorticity budget as 
\begin{equation}
\frac{\partial \zeta}{\partial t} = \nabla \times \left( \nabla^2 \frac{D \mathbf{u}}{Dt} \right) = \nabla^2 
\frac{D \zeta}{Dt} \nonumber, 
\end{equation}
where $\zeta = \nabla \times \mathbf{u}$ is the relative vorticity. 
Thus, if we add the RE-memory parametrization to a two-dimensional, non-divergent, unforced, inviscid equation for the vorticity budget, we obtain

\begin{eqnarray}
& &\frac{\partial \zeta}{\partial t} + \mathbf{u} \cdot \nabla \zeta + \beta v = \kappa \nabla^2 \frac{D \zeta}{Dt} \\
&\Rightarrow& (1 - \kappa \nabla^2) \left( \frac{\partial \zeta}{\partial t} + \mathbf{u} \cdot \nabla \zeta \right) + \beta v = 0, \nonumber
\end{eqnarray}
where $\beta$ is the planetary vorticity and, as before, $\kappa = \gamma (\Delta x)^2$ is a constant parameter and $-1 \leq \gamma \leq 1$. 
We linearize around a time-invariant zonal flow, $U$, and describe the flow using a stream function, $\zeta = \nabla^2 \Psi$. 
If we then make the ansatz $\Psi' = \Psi_0 e^{i(ik + ly - \omega t)}$, we arrive at a solution for the intrinsic phase speed for Rossby waves, 

\begin{equation}
c - U = - \frac{\beta k}{(1 + \kappa K^2) K^2}, \label{eq:phase_speed}
\end{equation}
but with a modification due to the RE-memory parametrization. 

This solution suggests faster westward phase speed when $\gamma < 0$ (and thus $\kappa < 0$), especially for short waves. 
As the mixing suppression across a jet is proportional to the ratio of the eddy kinetic energy and the squared intrinsic phase speed of eddies, $\mathrm{EKE}/(c-U)^2$ \citep{Ferrari2010}, the RE-memory parametrization could thus increase the mixing suppression by decreasing the intrinsic phase speed, especially for small wavenumbers. 
However, this effect also depends on how the RE-memory parametrization changes the KE of the flow (Eq. \ref{eq:ke_tend}).

\section{Relating the RE parameterization to the discovered equation of subfilter stress}

We found that the first term in the discovered equation \eqref{eqn:discovered_equation} can be related to the nonlinear gradient model (NGM), which has a tight connection with the RE stresses \cite{Anstey2017}. The NGM model of the second order is given by:
\begin{equation}
    \mathcal{S} = \alpha \nabla \cdot (\nabla \mathbf{u} \nabla \mathbf{u}^T), \label{eq:NGM2}
\end{equation}
and can be decomposed into two parts as follows:
\begin{equation}
     \nabla \cdot (\nabla \mathbf{u} \nabla \mathbf{u}^T) = \left( \nabla \mathbf{u} \nabla \right)^T \nabla \mathbf{u}^T + \nabla \mathbf{u} \nabla (\underbrace{\nabla \cdot \mathbf{u}}_{\approx 0}),
\label{eq:NGM2_decomposition}
\end{equation}
where the second term involves the horizontal divergence and is expected to be small, but may introduce small-scale noise.
The first term in the NGM model (Eq. \eqref{eq:NGM2_decomposition}) is the same as the leading term in the discovered equation (Eq. \eqref{eqn:discovered_equation}). Below, we give the specific implementation of how we compute the discovered operator in tensor notation:
\begin{gather}
    \left( \left( \nabla \mathbf{u} \nabla \right)^T \nabla \mathbf{u}^T \right)_i = 
    \frac{\partial^2 u_i}{\partial x_j \partial x_k}\frac{\partial u_j}{\partial x_k}, \label{eq:NGM2_tensor}
\end{gather}
where we assume the summation over the repeating indices.

\begin{figure}[h!]
    \centering
    \includegraphics[width=1.0\linewidth]{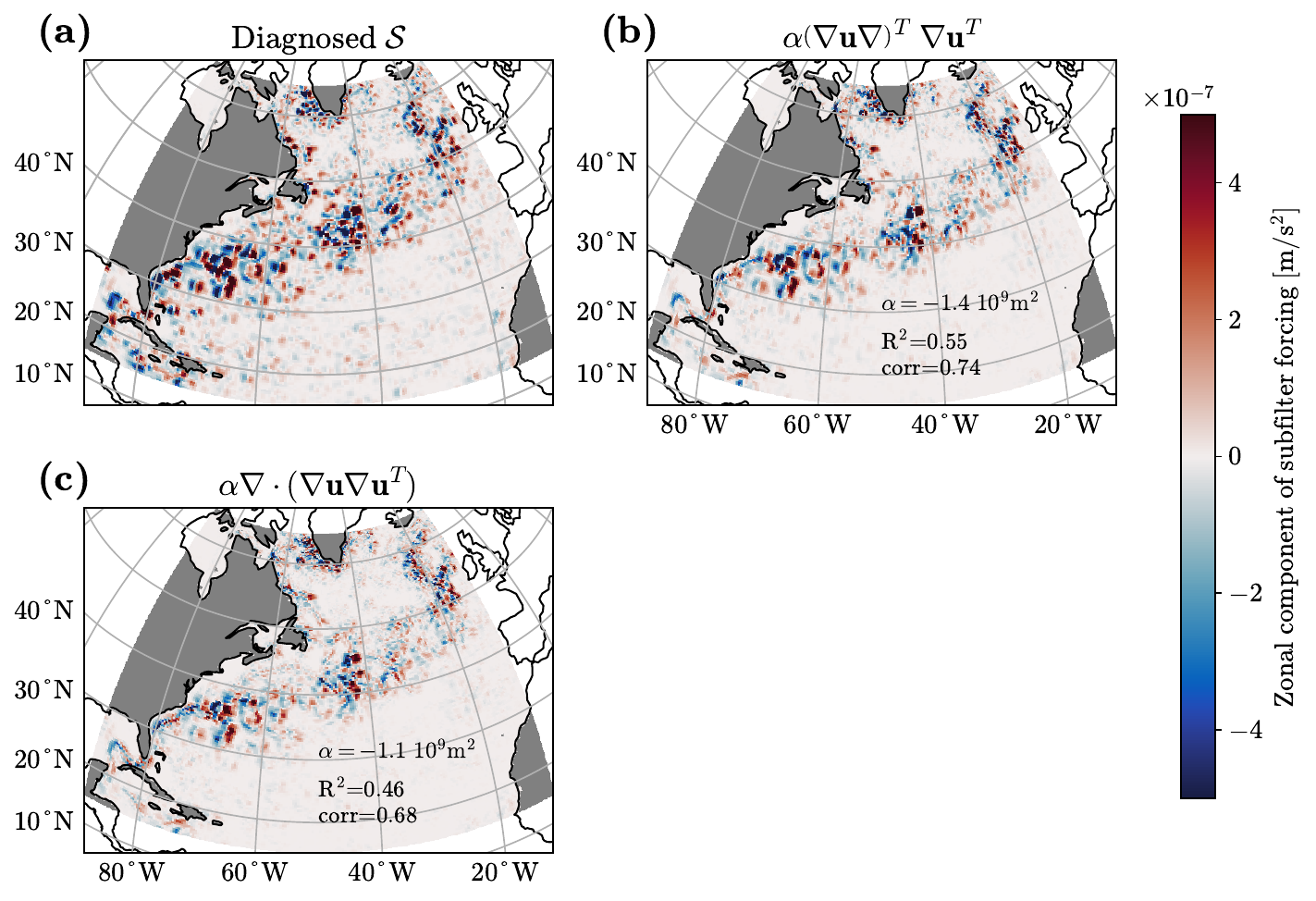}
    \caption{Zonal components of diagnosed (a) and predicted (b,c) subfilter forcing in the North Atlantic region at depth 5m, 5-day averaged. Regression coefficients ($\alpha$), coefficient of determination ($\mathrm{R}^2$), and pattern correlation (corr) are provided in the figure.}
    \label{fig:equations}
\end{figure}

In figure \ref{fig:equations} we compare the full NGM model (Eq. \eqref{eq:NGM2}) to the leading term in the discovered equation (Eq. \eqref{eq:NGM2_tensor}) on the dataset described above. The discovered equation predicts the diagnosed subfilter forcing with higher pattern correlation and higher coefficient of determination compared to the NGM model, while demonstrating smoother spatial patterns. We explain a more accurate prediction as a result of eliminating the term that depends on horizontal divergence, which introduces small-scale noise.

The second term in the discovered equation (Eq. \eqref{eqn:discovered_equation}), which is $\nabla \mathbf{u}^T \nabla^2 \mathbf{u}$, can be found in a decomposition of the tensor $\nabla \mathbf{u}^T \nabla \mathbf{u}$, which is proportional to the RE-AZ stress tensor (Eq. \eqref{eqn:AZ2017}) up to the isotropic stress:
\begin{equation}
    \nabla (\nabla \mathbf{u}^T \nabla \mathbf{u}) = \left( \nabla \mathbf{u}^T \nabla \right)^T \nabla \mathbf{u} + (\nabla \mathbf{u}^T \nabla^2 \mathbf{u}).
\end{equation}
However, the detailed analysis of the discovered term ($\nabla \mathbf{u}^T \nabla^2 \mathbf{u}$) goes beyond the scope of this paper.


\section{Summary.}

We have discussed the need for turbulence ocean closures in general for multiscale fluid problems with a focus on global ocean models. 
We also discussed the role of data-driven methods and theory in exploring possible alternatives to derive skillful predictions of ocean eddy stresses. 
In particular, we focused on diagnosing eddy stresses from a high-resolution simulation and exploring the skill of two possible closures. 
Namely, one closure that is based on an equation that is learned from data, and one closure that relates to a previously proposed closure based on Rivlin-Ericksen stresses. 
We explored and compared the properties of the closures, in terms of their ability to predict realistic spatial patterns of the eddy stresses and their contributions to the local and global momentum and energy cycle in the ocean.
We have demonstrated that there are connections with existing analytical closures, particularly with the nonlinear gradient model and the full Rivlin-Ericksen stress. 
However, we also highlight some discrepancies and the need for continued work in this area by combining numerics, analysis, and data-driven methods. 
We note that the explicit time-dependence of the RE-stress was not identified by the data-driven algorithm due to the insufficient temporal resolution in the existing dataset. However, based on the theoretical attributes of the term and its previous successful inclusion, we hope to explore this term further. In this note, we have not attempted to implement our closures into existing ocean models and leave this for future work.  

\section*{Acknowledgments.}
Pavel Perezhogin is a co-author of this study who co-led all aspects of this contribution (writing, theory, and computational analysis). This research received support through Schmidt Sciences, LLC, under the M$^2$LInES project. LZ also received support from NSF grant OCE 1912357 and NOAA CVP NA19OAR4310364. We thank all members of the M$^2$LInES team for their helpful discussions, advice, and support throughout this project. We are also grateful to Scott Bachman, James Anstey, and Joakim Kjellsson and Johanna Goldman for their help in conceptualizing aspects of this paper. This research was also supported in part through the NYU IT High Performance Computing resources, services, and staff expertise.


\bibliographystyle{siamplain}
\bibliography{Mendeley,agusample,bibliography}

@article{ross2022benchmarking,
  title={Benchmarking of machine learning ocean subgrid parameterizations in an idealized model},
  author={Ross, Andrew and Li, Ziwei and Perezhogin, Pavel and Fernandez-Granda, Carlos and Zanna, Laure},
  journal={Journal of Advances in Modeling Earth Systems},
  volume={15},
  number={1},
  pages={e2022MS003258},
  year={2023},
  publisher={Wiley Online Library},
  doi={https://doi.org/10.1029/2022MS003258}
}

@article{Wajsowicz1993,
    title = {{A consistent formulation of the anisotropic stress tensor for use in models of the large-scale ocean circulation}},
    year = {1993},
    journal = {Journal of Computational Physics},
    author = {Wajsowicz, RC},
    number = {2},
    month = {4},
    pages = {333--338},
    volume = {105},
    url = {http://linkinghub.elsevier.com/retrieve/pii/S002199918371079X http://www.sciencedirect.com.ezproxy.otago.ac.nz/science/article/pii/S002199918371079X},
    isbn = {0021-9991},
    doi = {10.1006/jcph.1993.1079},
    issn = {00219991}
}

@article{Anstey2017,
    title = {{A deformation-based parametrization of ocean mesoscale eddy reynolds stresses}},
    year = {2017},
    journal = {Ocean Modelling},
    author = {Anstey, James A. and Zanna, Laure},
    pages = {99--111},
    volume = {112},
    doi = {10.1016/j.ocemod.2017.02.004},
    issn = {14635003}
}

@article{Bachman2017,
    title = {{A scale-aware subgrid model for quasi-geostrophic turbulence}},
    year = {2017},
    journal = {Journal of Geophysical Research: Oceans},
    author = {Bachman, Scott D. and Fox-Kemper, Baylor and Pearson, Brodie},
    number = {2},
    month = {2},
    pages = {1529--1554},
    volume = {122},
    url = {http://doi.wiley.com/10.1002/2016JC012265},
    doi = {10.1002/2016JC012265},
    issn = {21699275}
}

@article{Smith2002,
    title = {{Anisotropic horizontal viscosity for ocean models}},
    year = {2002},
    journal = {Ocean Modelling},
    author = {Smith, Richard D. and McWilliams, James C.},
    number = {2},
    pages = {129--156},
    volume = {5},
    doi = {10.1016/S1463-5003(02)00016-1},
    issn = {14635003}
}

@article{Griffies2000,
    title = {{Biharmonic Friction with a Smagorinsky-Like Viscosity for Use in Large-Scale Eddy-Permitting Ocean Models}},
    year = {2000},
    journal = {Monthly Weather Review},
    author = {Griffies, Stephen M and Hallberg, Robert W},
    number = {8},
    pages = {2935--2946},
    volume = {128},
    url = {http://dx.doi.org/10.1175/1520-0493(2000)128<2935:BFWASL>2.0.CO;2},
    isbn = {0027-0644},
    doi = {10.1175/1520-0493(2000)128<2935:BFWASL>2.0.CO;2},
    issn = {0027-0644}
}

@article{Brankart2013,
    title = {{Impact of uncertainties in the horizontal density gradient upon low resolution global ocean modelling}},
    year = {2013},
    journal = {Ocean Modelling},
    author = {Brankart, Jean Michel},
    pages = {64--76},
    volume = {66},
    isbn = {1463-5003},
    doi = {10.1016/j.ocemod.2013.02.004},
    issn = {14635003}
}

@article{Hecht2008,
    title = {{Implementation of the LANS-{$\alpha$} turbulence model in a primitive equation ocean model}},
    year = {2008},
    journal = {Journal of Computational Physics},
    author = {Hecht, Matthew W. and Holm, Darryl D. and Petersen, Mark R. and Wingate, Beth A.},
    number = {11},
    pages = {5691--5716},
    volume = {227},
    isbn = {bibcode:2008JCoPh.227.5691H},
    doi = {10.1016/j.jcp.2008.02.018},
    issn = {00219991},
    arxivId = {physics/0703195}
}

@article{Jansen2014,
    title = {{Parameterizing subgrid-scale eddy effects using energetically consistent backscatter}},
    year = {2014},
    journal = {Ocean Modelling},
    author = {Jansen, Malte F. and Held, Isaac M.},
    pages = {36--48},
    volume = {80},
    doi = {10.1016/j.ocemod.2014.06.002},
    issn = {14635003}
}

@article{Zanna2017,
    title = {{Scale-aware deterministic and stochastic parametrizations of eddy-mean flow interaction}},
    year = {2017},
    journal = {Ocean Modelling},
    author = {Zanna, Laure and Porta Mana, PierGianLuca and Anstey, James and David, Tomos and Bolton, Thomas},
    pages = {66--80},
    volume = {111},
    url = {http://www.sciencedirect.com/science/article/pii/S1463500317300100},
    doi = {10.1016/j.ocemod.2017.01.004},
    issn = {14635003}
}

@article{Rivlin1955,
    title = {{Stress-Deformation Relations for Isotropic Materials}},
    year = {1955},
    journal = {Indiana Univ. Math. J.},
    author = {Rivlin, R. and Ericksen, J.},
    number = {2},
    pages = {323--425},
    volume = {4},
    issn = {0022-2518}
}

@article{Ferrari2010,
    title = {{Suppression of Eddy Diffusivity across Jets in the Southern Ocean}},
    year = {2010},
    journal = {Journal of Physical Oceanography},
    author = {Ferrari, Raffaele and Nikurashin, Maxim},
    number = {7},
    pages = {1501--1519},
    volume = {40},
    url = {http://journals.ametsoc.org/doi/abs/10.1175/2010JPO4278.1},
    isbn = {0022-3670},
    doi = {10.1175/2010JPO4278.1},
    issn = {0022-3670}
}

@article{Holm2005,
    title = {{The LANS-{$\alpha$} model for computing turbulence}},
    year = {2005},
    journal = {Los Alamos Science},
    author = {Holm, Darryl D and Jeffery, C. and Kurien, S. and Livescu, D. and Taylor, M.A. and Wingate, Beth A.},
    pages = {152--171},
    volume = {29}
}

@article{Bachman2018,
    title = {{The relationship between a deformation-based eddy parameterization and the LANS-{$\alpha$} turbulence model}},
    year = {2018},
    journal = {Ocean Modelling},
    author = {Bachman, Scott D. and Anstey, James A. and Zanna, Laure},
    pages = {56--62},
    volume = {126},
    doi = {10.1016/j.ocemod.2018.04.007},
    issn = {14635003}
}

@article{PortaMana2014,
    title = {{Toward a stochastic parameterization of ocean mesoscale eddies}},
    year = {2014},
    journal = {Ocean Modelling},
    author = {Porta Mana, PierGianLuca and Zanna, Laure},
    pages = {1--20},
    volume = {79},
    doi = {10.1016/j.ocemod.2014.04.002}
}

@article{Hallberg2013,
    title = {{Using a resolution function to regulate parameterizations of oceanic mesoscale eddy effects}},
    year = {2013},
    journal = {Ocean Modelling},
    author = {Hallberg, Robert},
    month = {12},
    pages = {92--103},
    volume = {72},
    url = {https://linkinghub.elsevier.com/retrieve/pii/S1463500313001601},
    doi = {10.1016/j.ocemod.2013.08.007},
    issn = {14635003}
}

@article{guillaumin2021stochastic,
  title={Stochastic-deep learning parameterization of ocean momentum forcing},
  author={Guillaumin, Arthur P and Zanna, Laure},
  journal={Journal of Advances in Modeling Earth Systems},
  volume={13},
  number={9},
  pages={e2021MS002534},
  year={2021},
  publisher={Wiley Online Library},
  doi={https://doi.org/10.1029/2021MS002534}
}

@incollection{christensen2022parametrization,
  doi = {10.1093/acrefore/9780190228620.013.826},
  year = {2022},
  ISBN = {9780190228620},
  booktitle = {Oxford Research Encyclopedia of Climate Science},
  publisher = {Oxford University Press},
  author = {Hannah Christensen and Laure Zanna},
  title = {{Parametrization in Weather and Climate Models}}
}

@article{griffies2015impacts,
  title={Impacts on ocean heat from transient mesoscale eddies in a hierarchy of climate models},
  author={Griffies, Stephen M and Winton, Michael and Anderson, Whit G and Benson, Rusty and Delworth, Thomas L and Dufour, Carolina O and Dunne, John P and Goddard, Paul and Morrison, Adele K and Rosati, Anthony and others},
  journal={Journal of Climate},
  volume={28},
  number={3},
  pages={952--977},
  year={2015},
  publisher={American Meteorological Society},
  doi={https://doi.org/10.1175/JCLI-D-14-00353.1}
}

@article{smagorinsky1963general,
  title={General circulation experiments with the primitive equations: I. The basic experiment},
  author={Smagorinsky, Joseph},
  journal={Monthly weather review},
  volume={91},
  number={3},
  pages={99--164},
  year={1963},
  publisher={American Meteorological Society},
  doi={https://doi.org/10.1175/1520-0493(1963)091%3C0099:GCEWTP%3E2.3.CO;2}
}

@article{perezhogin2025generalizable,
  title={Generalizable neural-network parameterization of mesoscale eddies in idealized and global ocean models},
  author={Perezhogin, Pavel and Adcroft, Alistair and Zanna, Laure},
  journal={arXiv preprint arXiv:2505.08900},
  year={2025}
}

@misc{perezhogin_2025_dataset,
  author       = {Perezhogin, Pavel and
                  Adcroft, Alistair and
                  Zanna, Laure},
  title        = {{Generalizable neural-network parameterization of mesoscale eddies in idealized and global ocean models [Dataset]
                  }},
  year         = 2025,
  publisher    = {Zenodo},
  version      = {v6},
  doi          = {https://doi.org/10.5281/zenodo.16058005},
}

@string{AS = {Ann.~Stat.}}

@article{li2021,
  title={Data-driven model development for large-eddy simulation of turbulence using gene-expression programing},
  author={Li, Haochen and Zhao, Yaomin and Wang, Jianchun and Sandberg, Richard D},
  journal={Physics of Fluids},
  volume={33},
  number={12},
  pages={125127},
  year={2021},
  publisher={AIP Publishing LLC}
}

@misc{gplearn,
    author={Stephens, Trevor},
    title={{Genetic Programming in Python, with a scikit-learn inspired API}},
    year={2019},
    url={https://gplearn.readthedocs.io/en/stable},
    version={0.4.1}
}

@article{koza1994genetic,
  title={Genetic programming as a means for programming computers by natural selection},
  author={Koza, John R},
  journal={Statistics and computing},
  volume={4},
  number={2},
  pages={87--112},
  year={1994},
  publisher={Springer}
}

@article{Zanna2020,
author = {Zanna, Laure and Bolton, Thomas},
title = {Data-driven Equation Discovery of Ocean Mesoscale Closures},
journal = {Geophysical Research Letters},
pages = {e2020GL088376},
year = 2020,
doi={10.1029/2020GL088376},
}

@article{10.1093/mind/LIX.236.433,
    author = {Turing, Alan},
    title = "{I.—COMPUTING MACHINERY AND INTELLIGENCE}",
    journal = {Mind},
    volume = {LIX},
    number = {236},
    pages = {433-460},
    year = {1950},
    month = {10},
    issn = {0026-4423},
    doi = {10.1093/mind/LIX.236.433},
    url = {https://doi.org/10.1093/mind/LIX.236.433},
    eprint = {https://academic.oup.com/mind/article-pdf/LIX/236/433/30123314/lix-236-433.pdf},
}

@article{brunton2016sparse,
  title={Sparse identification of nonlinear dynamics with control (SINDYc)},
  author={Brunton, Steven L and Proctor, Joshua L and Kutz, J Nathan},
  journal={IFAC-PapersOnLine},
  volume={49},
  number={18},
  pages={710--715},
  year={2016},
  publisher={Elsevier}
}
\end{document}